\documentclass[12pt]{iopart}
\pdfoutput=1

\usepackage{mathptmx,courier,pifont}
\usepackage[scaled=0.92]{helvet}
\usepackage[T1]{fontenc}
\usepackage{textcomp}
\usepackage{mathastext}

\usepackage{epsfig}

\usepackage{graphicx}
\usepackage{mathtools,cuted}
\usepackage{dcolumn}
\usepackage{bm}
\usepackage{braket}
\usepackage{textgreek}
\usepackage{rotate}
\usepackage[table]{xcolor}
\usepackage{tabularx}
\usepackage{footmisc,booktabs}
\usepackage[referable]{threeparttablex}

\usepackage{letltxmacro,etoolbox,booktabs}

\LetLtxMacro{\originalcite}{\cite}
\def\tablecite#1#{%
  \def\pretablecite{#1}%
  \tableciteaux}
\def\tableciteaux#1{%
  \textsuperscript{\expandafter\originalcite\pretablecite{#1}}%
}
\AtBeginEnvironment{table}{\let\cite\tablecite}

\definecolor{capri}{rgb}{0.0, 0.75, 1.0}
\definecolor{cornflowerblue}{rgb}{0.39, 0.58, 0.93}
\definecolor{spirodiscoball}{rgb}{0.06, 0.75, 0.99}
\definecolor{pear}{rgb}{0.82, 0.89, 0.19}

\usepackage{changepage}

\usepackage{hyperref}

\usepackage[T1]{fontenc}
\usepackage{calrsfs}
\DeclareMathAlphabet{\mathdutchcal}{U}{dutchcal}{m}{n}

\usepackage{upgreek}

\begin{document}

\title[Emergence of Cluster Formation in Light Nuclei]{Emergence of Cluster Formation in Light Nuclei}

\author{Jos\'e Nicol\'as Orce$^{1,2}$ and Manfred Jason Jaftha$^1$}
\address{$^1$ Department of Physics \& Astronomy, University of the Western Cape, P/B X17, Bellville ZA-7535, South Africa}
\address{$^2$ National Institute for Theoretical and Computational Sciences (NITheCS), Stellenbosch, South Africa}
\ead{jnorce@uwc.ac.za; nico.orce@cern.ch; https://nuclear.uwc.ac.za}


\begin{abstract}

Spherical harmonics form a complete orthonormal basis, allowing any function on the sphere to be expanded.
The nuclear shape of a given eigenstate can thus be described within Bohr’s quasi-molecular model by a coordinate transformation
from a randomly oriented ellipsoid in space to a coordinate system aligned with the ellipsoid’s principal axes.
This transformation (Eq.~\eqref{eq:sh1}) is characterized by three Euler angles and two deformation parameters,
$\beta$ (quadrupole) and $\gamma$ (triaxiality), but does not uniquely define the nuclear shape;
rotational averaging over equivalent orientations is expected to yield a diffuse nuclear shape.
Rotational invariance under $\beta$ and $\gamma$ is achieved using  three transformation operators,
which define a new coordinate system aligned with a single intrinsic configuration (Eq.~\eqref{eq:sh2}).
Here we show that the nonunique coordinate system of Eq.~\eqref{eq:sh1} with $\beta$ and $\gamma$ deformation parameters
extracted from experimental electric-quadrupole matrix elements unexpectedly yields the most probable nuclear shape.
In fact, only then does cluster formation spatially emerge, reproducing the characteristic bowling-pin-like shapes of $^{10}$B and $^{20}$Ne,
consistent with modern nuclear theory.
Both coordinate systems generally exhibit the same shape features for heavier deformed nuclei, where
substantial triaxial deformation is empirically observed. However, the approach based on Eq.~\eqref{eq:sh1},
using empirical $\beta$ and $\gamma$ values, provides deeper insight by capturing the superposition of
multiple intrinsic configurations that collectively form the nuclear state. This, in turn, offers a physical interpretation of triaxiality.

\end{abstract}

\noindent{\it Keywords}: spherical harmonics, spectroscopic quadrupole moment, triaxiality, $\alpha$ clusters, quasi-molecular model, electric quadrupole matrix elements
\maketitle

\section{Motivation}

This work is motivated by novel research at the Large Hadron Collider ({\scshape LHC}) at {\scshape CERN} probing nuclear geometry via light-ion collisions ({\scshape LIC}).
Here, nuclear shapes --- including $\alpha$-cluster configurations, deviations from
axial symmetry~\cite{brewer2021,giacalone2024,jia2024imaging,wang2024deformation,liu2025directly} and shape fluctuations~\cite{dimri2023impact} ---  
may be inferred from the  overlap region or centrality of the two colliding nuclei~\cite{aaij2022centrality}.
One of the most crucial shapes concerns the nucleus $^{20}$Ne~\cite{ebran2012atomic},
where a bowling pin-like cluster formation is expected from an increased incoherent cross section
relative to a spherical shape~\cite{mantysaari2023multiscale}.

\begin{figure}[!ht]
\begin{center}
\includegraphics[width=6cm,height=4.5cm,angle=-0]{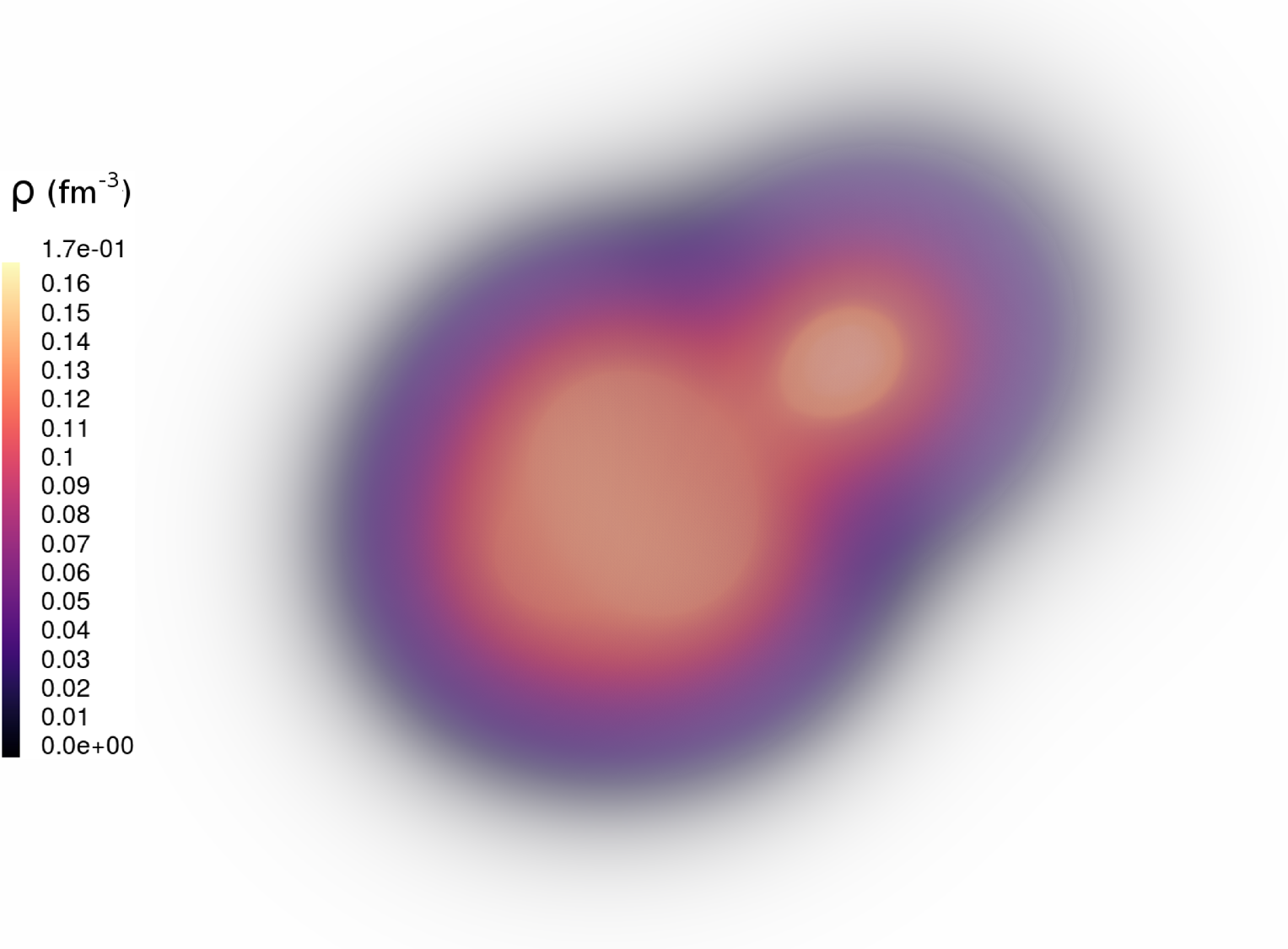}
\caption{Characteristic  intrinsic nucleon density of the 2$_1^+$ state in $^{20}$Ne obtained with the {\scshape MR-EDF} model. This is the {\scshape 3D} version of Fig. 7 in Ref.~\cite{mehl2025large}, courtesy of J.-P. Ebran.}
\label{fig:20Ne}
\end{center}
\end{figure}

Such a bowling-pin shape for the ground state of $^{20}$Ne has long been predicted by antisymmetrized molecular dynamics ({\scshape AMD}) calculations~\cite{kanada1995clustering},
along with their modern extensions~\cite{myo2025shell},  as arising from many-body correlations of nucleons described by antisymmetrized products of Gaussian wave packets~\cite{kanada2012antisymmetrized}.
An independent mean-field approach
based on energy-density functionals ({\scshape EDF}) also predicts a  bowling-pin intrinsic deformation using the relativistic  {\scshape DD-ME2} functional~\cite{ebran2012atomic},
whose deep single-nucleon potentials favor more pronounced cluster structures.
Similar multireference relativistic-{\scshape EDF} calculations ({\scshape MR-EDF}), based on the relativistic Hartree-Bogoliubov ({\scshape RHB}) model
and using the relativistic {\scshape DD-PC1} functional, further predict a bowling-pin-like shape (shown in Fig.~\ref{fig:20Ne})
for the first excited $J^{\pi}$$=$$2^+_1$ state in $^{20}$Ne~\cite{mehl2025large}.

Analogous conclusions are  reached using other computationally-demanding theoretical approaches~\cite{funaki2014alpha,Neff,nb1m-cyr3},
including
modern {\it ab initio} methods such as the symmetry-adapted no-core shell model ({\scshape SA-NCSM}) and the projected generator coordinate method ({\scshape PGCM})~\cite{giacalone2025exploiting},
all of which consistently predict a bowling-pin-like shape for the ground state of $^{20}$Ne.
A more planar geometry is predicted for $^{20}$Ne by the $\alpha$-cluster model~\cite{sosin2016alpha}, in which the central $\alpha$ cluster
is more diffused than the peripheral ones. Here, the appearance of $\alpha$ clusters arises from additional spin-polarization and symmetry-energy terms
in the expansion of the equation of state (EoS) around saturation point.
Recent review articles provide a comprehensive overview of the history and recent developments of $\alpha$
clustering in $^{20}$Ne and other light nuclei~\cite{freer2018microscopic,lombardo2023clusters}.

\section{Quasi-Molecular Model Revisited}

Despite general agreement, it was not until recently that modern nuclear theory could account for the large spectroscopic quadrupole moment measured in $^{20}$Ne~\cite{mehl2025large};
which quantifies the extent to which the charge distribution in the laboratory frame assumes an ellipsoidal shape.
To this end, state-of-the-art calculations using nuclear lattice effective field theory ({\scshape NLEFT}) incorporate $\alpha$-cluster correlations within the minimal nuclear interaction~\cite{giacalone2025exploiting}.
The emergent geometry and intrinsic cluster structure are determined by the so-called pinhole algorithm that pins down the center-of-mass of $^{20}$Ne,
thereby enabling the reproduction of both the spectroscopic quadrupole moment and the characteristic bowling-pin shape.

Recent beyond Skyrme–Hartree–Fock ({\scshape SHF}) calculations with tensor forces using the  {\scshape SAM}i interaction~\cite{roca2012new} also reproduce
the large spectroscopic quadrupole moment in $^{20}$Ne~\cite{nb1m-cyr3}, highlighting the role of tensor forces in
enhancing clustering along the $z$-symmetry axis.
A similar effect is observed by {\scshape NLEFT} calculations when aligning cluster configurations to the $z$-direction~\cite{giacalone2025exploiting}.
By contrast, most {\it ab initio} methods omit cluster correlations and tend to underestimate electric quadrupole ({\scshape E2}) properties, including reduced transition probabilities (or {\scshape B(E2)} values) and spectroscopic quadrupole moments~\cite{ngwetsheni2025extremely,mehl2025large}.

Beyond strictly microscopic approaches, nuclear structure physics has undoubtedly benefit from the seminal work of Aage Bohr in 1952, `The coupling of nuclear surface oscillations to the motion of individual nucleons'~\cite{bohr1952coupling},
which led to both vibrational and rotational interpretations of nuclear collectivity~\cite{bohr1998nuclear}.
Specifically, the angular part of the wave function can be represented by spherical harmonics  $Y_\ell^m(\theta,\phi)$~\cite{marquis1825traite},
which form a complete orthonormal basis for all square-integrable functions  on the surface of a sphere.
In a coordinate system where the axes coincide with the principal axes of the ellipsoid, the nuclear shape of an eigenstate can
be given in polar spherical coordinates,
\begin{equation}
R(\theta,\phi) = R_0 \left(1+ \sum_{m} \alpha_{2,m} Y_2^m(\theta,\phi) \right),
\end{equation}
where only $Y_2^m$ harmonics with $m$$=$$0,\pm2$ are considered,
$R_0$$=$$1.2 A^{1/3}$ fm is the spherical radius, and $\alpha_{2,m}$ are the coordinates describing the deformation of the nuclear surface.
Structural effects are, in principle, neglected.
Conveniently, $\alpha_{2,m}$ can be substituted by $\beta$ (quadrupole) and $\gamma$ (triaxiality) deformation parameters~\cite{hill1953nuclear,verney2025history},
\begin{eqnarray}
\alpha_{2,0} &=& \beta~\cos\gamma \label{eq:alphabeta1} \\
\alpha_{2,2}=\alpha_{2,-2} &=& \frac{1}{\sqrt{2}} ~\beta~\sin\gamma \label{eq:alphabeta2},
\label{eq:alphhabeta1}
\end{eqnarray}
where $\alpha_{2,0}$ indicates the stretching of the $z$-axis
with respect to the $x$- and $y$-axes,
and $\alpha_{2,2}$ the difference in length between the $x$- and $y$-axes, with $\alpha_{2,2}=0$ for an axially symmetric prolate shape.
The nuclear shape is then characterized by $~Y_2^0$, $Y_2^2$ and $Y_2^{-2}$,
\begin{equation}
  \begin{split}
R(\theta,\phi) &= R_0 \Biggl( 1+ \beta \sqrt{\frac{5}{16\pi}} ~\biggl[ \cos\gamma~ (3\cos^2\theta -1)  \\ &+ \sqrt{3} \sin\gamma \sin^2\theta \cos(2\phi) \biggr] \Biggr). \label{eq:sh1}
\end{split}
\end{equation}

It is important to note here that Eq.~\eqref{eq:sh1} does not uniquely determine the nuclear shape in the new coordinate system aligned with the principal axes of the ellipsoid~\cite{bohr1952coupling}.
There are 24 distinct choices of right-handed coordinate systems with axes aligned along the ellipsoidal axes
--- and infinitely many for an axially-symmetric shape --- that  may affect the symmetry properties of the wave function.
Empirical~\cite{triaxialityElenaNico,triaxiality2ElenaNico} and theoretical~\cite{moller2006global,casten2000nuclear,otsuka2025prevailing}
studies indicate, however, that most nuclei are quadrupole deformed and exhibit axially asymmetric (or triaxial) shapes.
Invariance under $\beta$ and $\gamma$ can then be expressed in terms of three transformation operators~\cite{bohr1952coupling}
\begin{eqnarray}
R_1 &=& R_0 \biggl( 1+ \beta \sqrt{\frac{5}{16\pi} \bigl[ -\cos\gamma+\sqrt{3} \sin\gamma} \bigr] \biggr) \nonumber  \\
R_2 &=& R_0 \biggl( 1+ \beta \sqrt{\frac{5}{16\pi} \bigl[ -\cos\gamma-\sqrt{3} \sin\gamma} \bigr] \biggr)  \nonumber  \\
R_3 &=& R_0 \biggl( 1+  \beta \sqrt{\frac{5}{16\pi}} ~2\cos\gamma \biggr),
\label{eq:eulersimple}
\end{eqnarray}
yielding the general expression,
\begin{equation}
  \begin{split}
R_k(\theta,\phi) &= R_0 \Biggl( 1+ \beta \sqrt{\frac{5}{4\pi}} \cos\left(\gamma - \frac{2{\pi}k}{3} \right) \Biggr)~\mbox{for k=1,2,3}, \label{eq:sh2}
\end{split}
\end{equation}
which map the coordinate system onto a single intrinsic configuration.

\begin{figure*}[!htbp]
\begin{center}
\includegraphics[width=5.9cm,height=5.9cm,angle=-0]{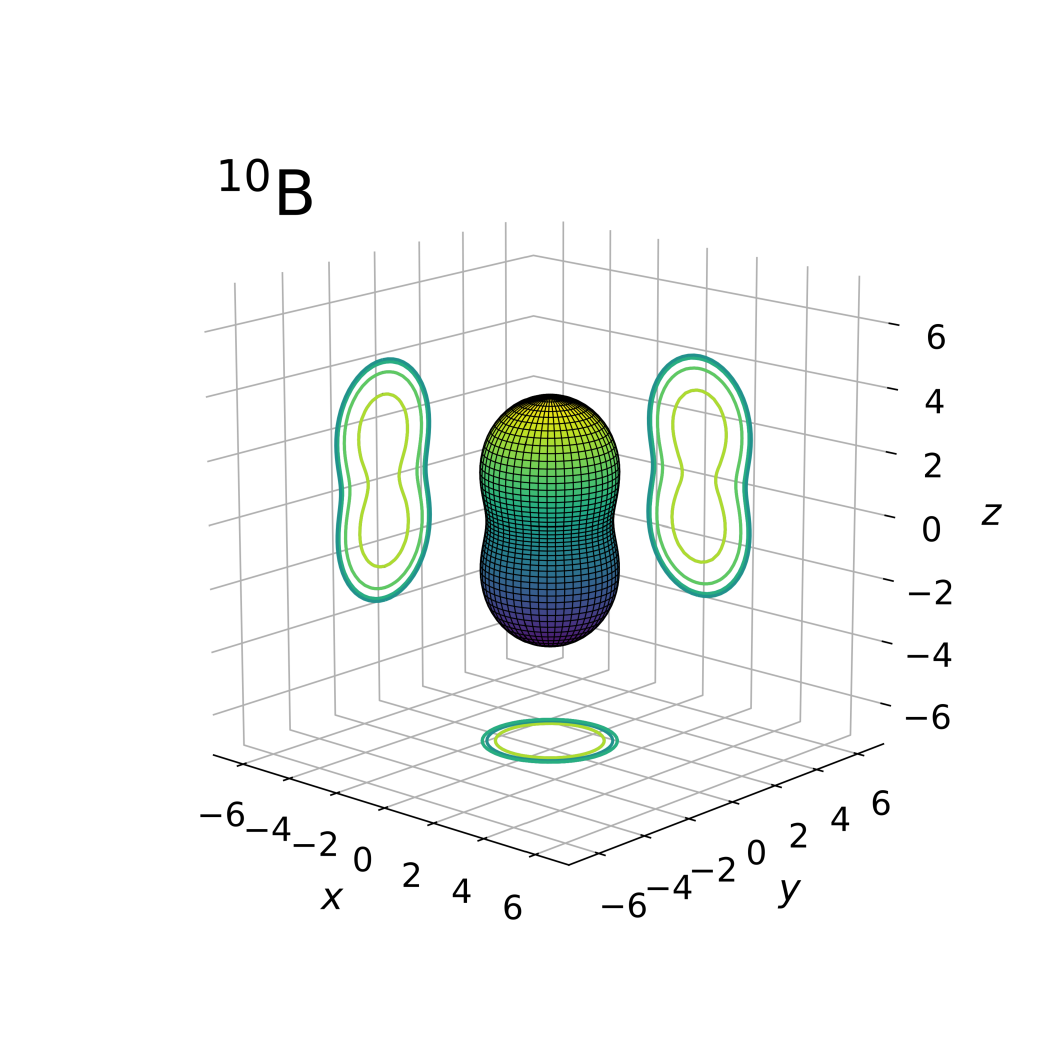} \hspace{-0.4cm}
\includegraphics[width=5.1cm,height=5.1cm,angle=-0]{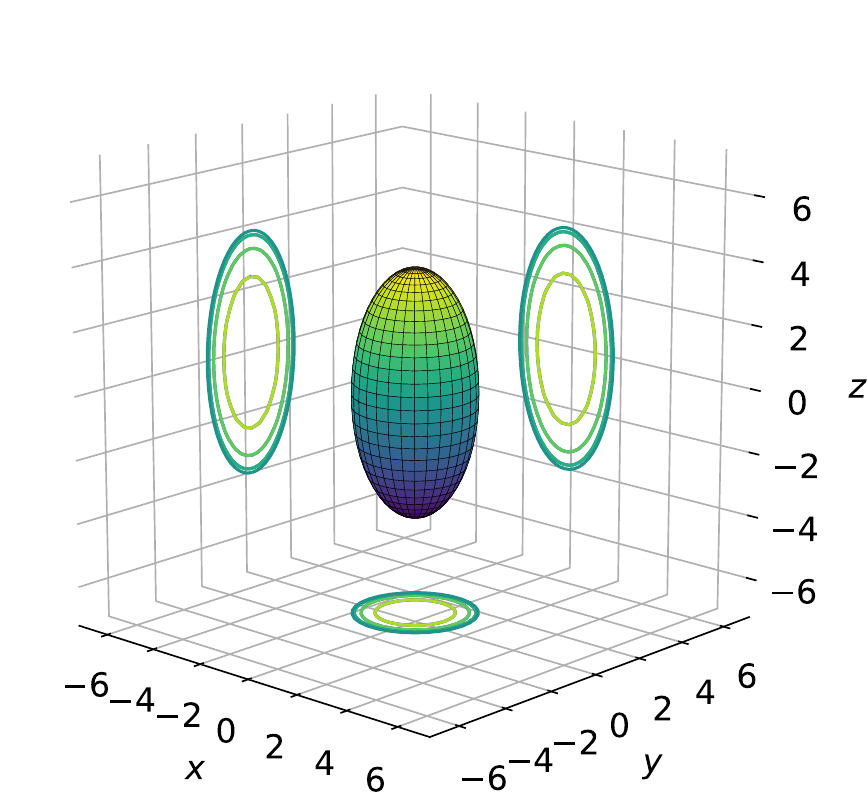}  \\ \vspace{-0.7cm}
\includegraphics[width=5.9cm,height=5.9cm,angle=-0]{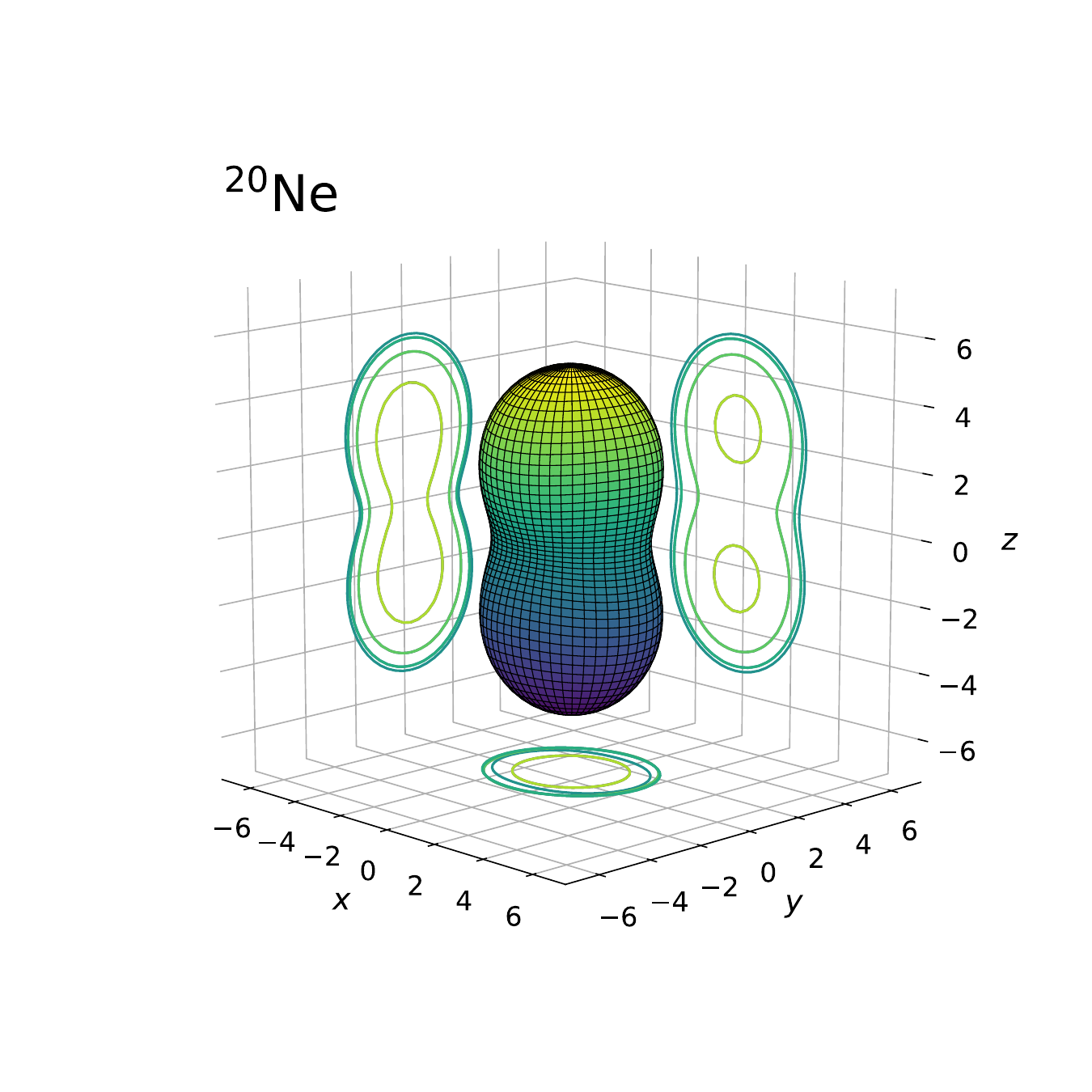} \hspace{-0.5cm}
\includegraphics[width=5.2cm,height=5.2cm,angle=-0]{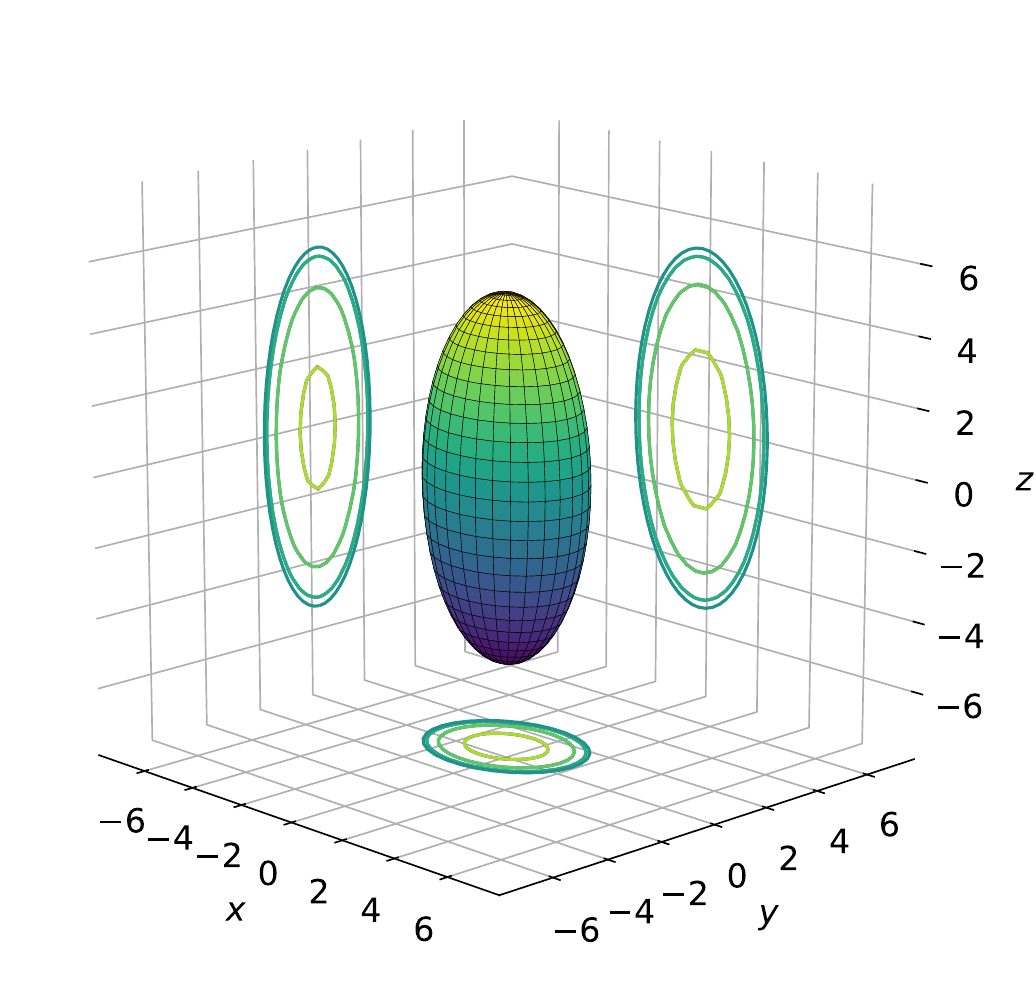}  \\ \vspace{-0.7cm}
\includegraphics[width=6.2cm,height=6.2cm,angle=-0]{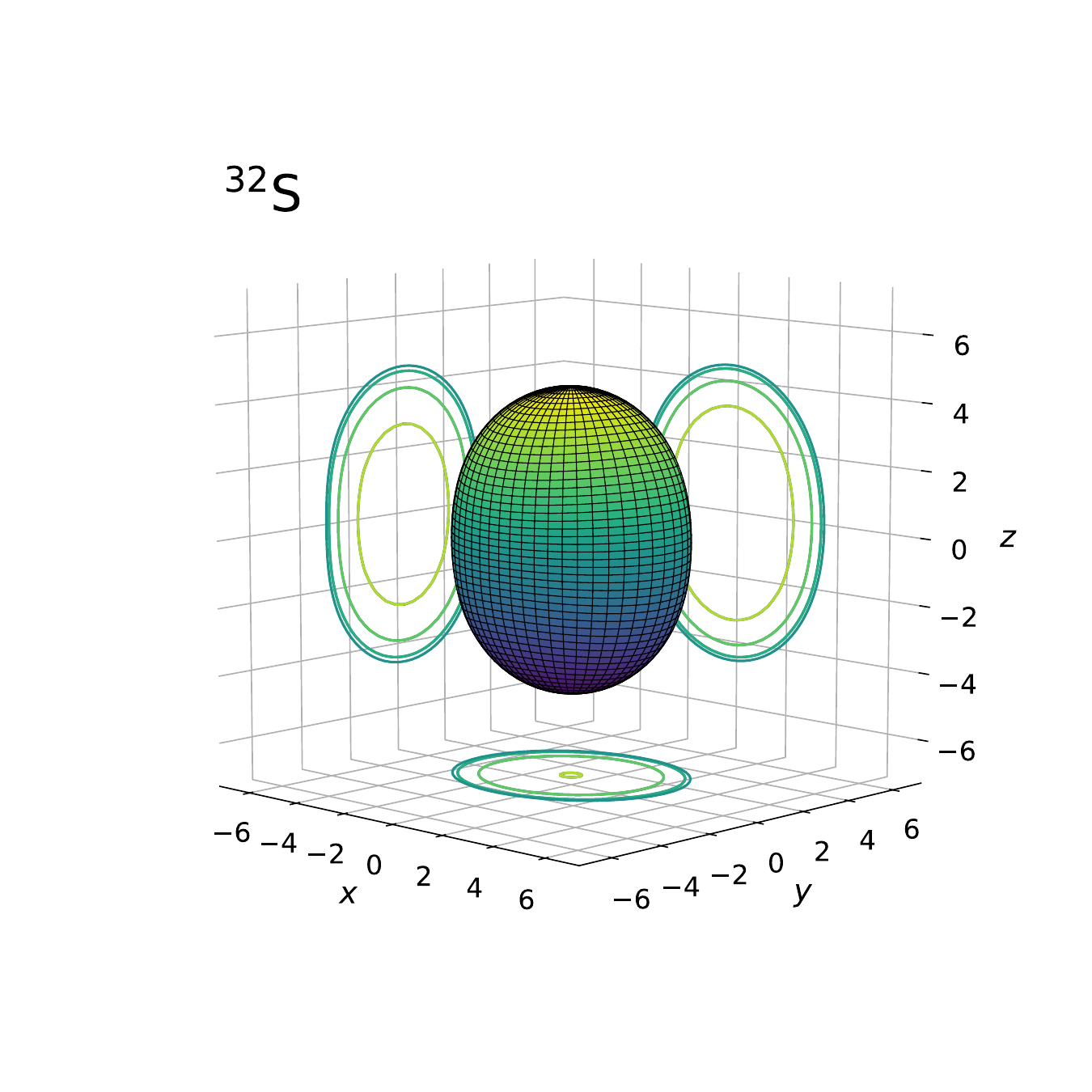} \hspace{-0.5cm}
\includegraphics[width=5.5cm,height=5.5cm,angle=-0]{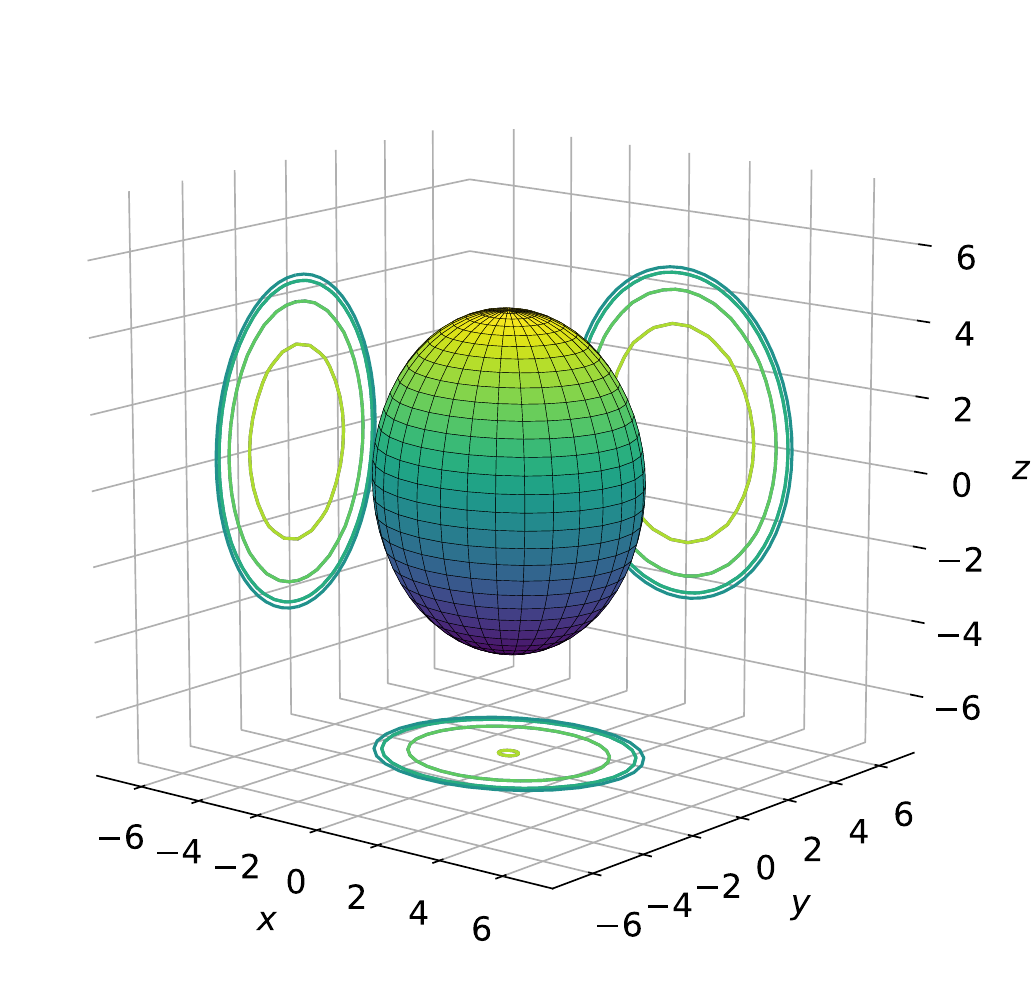}  \\ \vspace{-0.7cm}
\includegraphics[width=6.2cm,height=6.2cm,angle=-0]{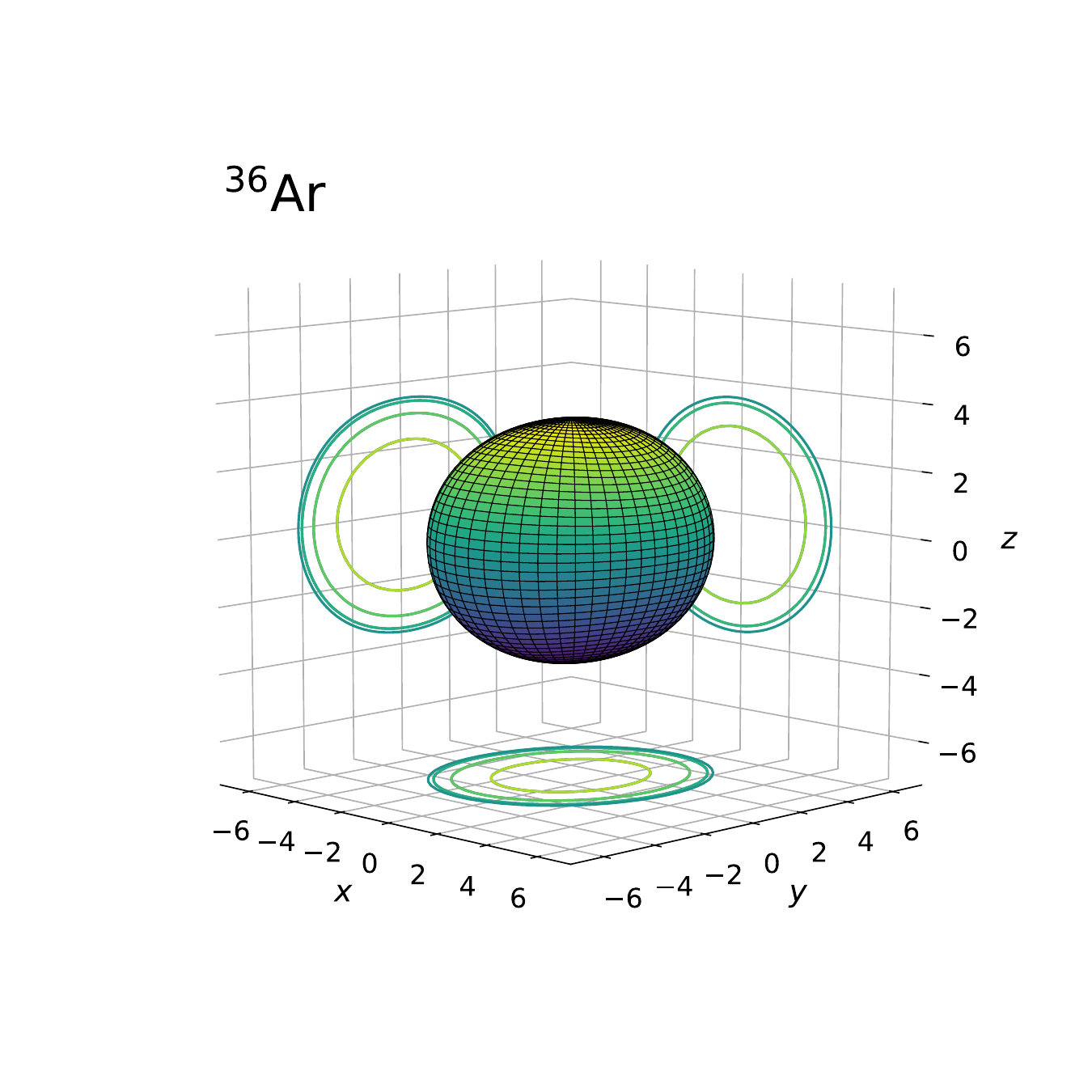} \hspace{-0.5cm}
\includegraphics[width=5.5cm,height=5.5cm,angle=-0]{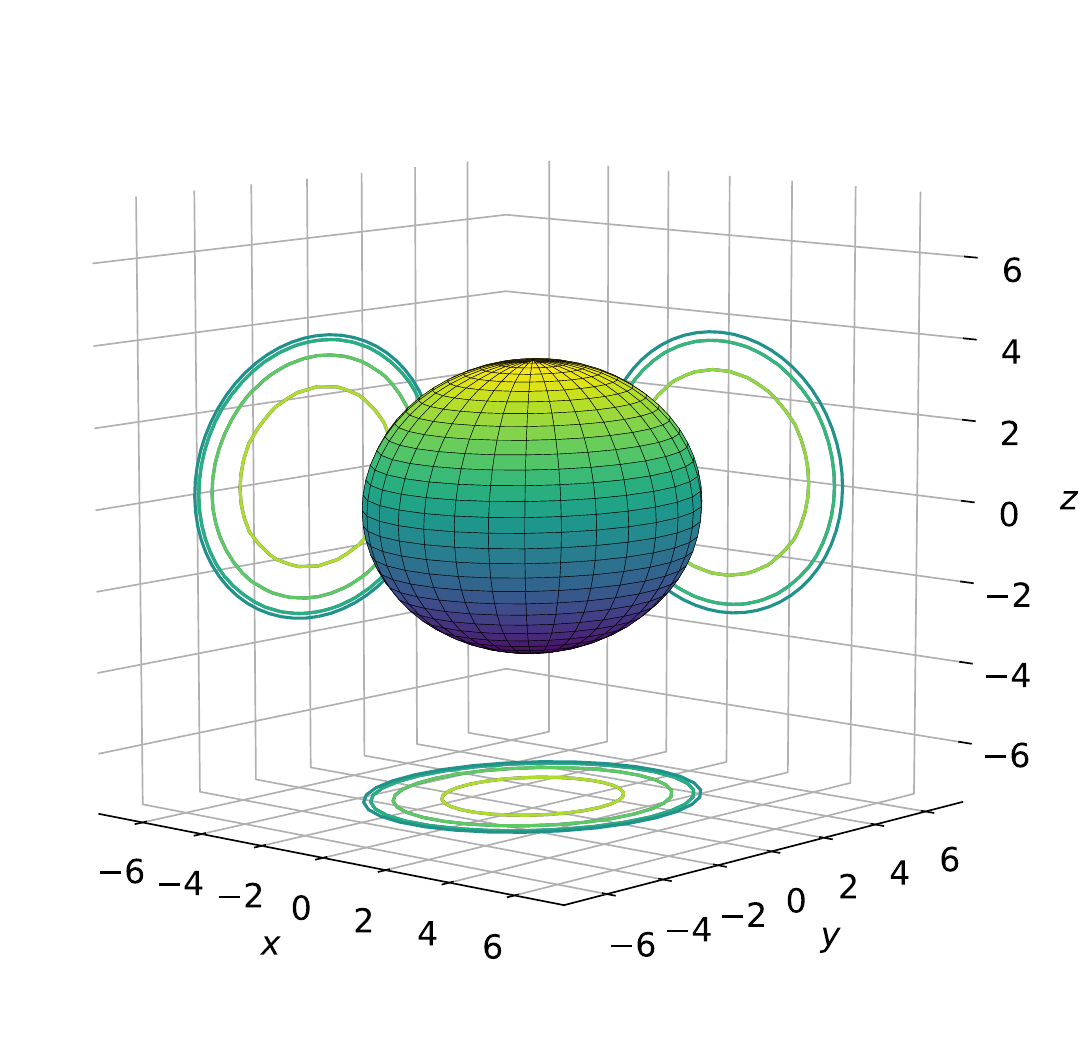}
\caption{Evolution of triaxial shapes for the 4$n$ self-conjugate nuclei $^{10}$B (top), $^{20}$Ne (upper middle), $^{32}$S (lower middle)
and $^{36}$Ar (bottom), obtained using Eq.~\eqref{eq:sh1} (left) and Eqs.~\eqref{eq:sh2} (right).
The axes units are fm~\cite{manfred}.}
\label{shape20Ne}
\end{center}
\end{figure*}

\section{Results and Discussion}

Nuclear shapes obtained using Eqs.~\eqref{eq:sh1} and \eqref{eq:sh2} with empirical $\beta$ and $\gamma$ deformations
are shown in the left and right panels of Fig.~\ref{shape20Ne},
respectively, for  $^{10}$B (top), $^{20}$Ne (upper middle), $^{32}$S (lower middle) and $^{36}$Ar (bottom).
Here, $\beta$ is determined from the measured spectroscopic quadrupole moment~\cite{bohr1998nuclear} and $\gamma$ from the empirical triaxial rotor model~\cite{triaxialityElenaNico,triaxiality2ElenaNico}.
Empirical triaxialities are  in overall agreement with the ones measured model independently using
rotational invariants of zero-coupled products of the  {\scshape$\hat{E2}$} electric quadrupole operator~\cite{hadynska2016superdeformed,kisyov2022structure,wu1996quadrupole},
which are also built by combining spherical harmonics or ranked-2 tensors derived from them~\cite{kumar1972intrinsic,cline1986heavy}.

More pronounced shape features are observed in light nuclei using Eq.~\eqref{eq:sh1}, i.e.,
without mapping the coordinate system (Eqs.~\ref{eq:eulersimple}) onto a single intrinsic configuration (Eq.~\eqref{eq:sh2}).
As shown in the right panels of Fig.~\ref{shape20Ne}, while Eq.~\eqref{eq:sh2} give rise to smooth prolate or rugby-ball shapes for $^{10}$B (top) and $^{20}$Ne (upper middle),
the corresponding nuclear shapes obtained by Eq.~\eqref{eq:sh1} in the left panels  of Fig.~\ref{shape20Ne} show the characteristic bowling-pin- or rather peanut-like shapes.
Shape parameters of ($\beta$$=$$0.80$~\cite{nesbet1970atomic}, $\gamma$$=$$0^{\circ}$) and ($\beta$$=$$0.96$~\cite{mehl2025large}, $\gamma$$=$$9.2^{\circ}$~\cite{triaxialityElenaNico})
were used for $^{10}$B and $^{20}$Ne, respectively.
Although $\gamma$ is not empirically known for $^{10}$B, the large spectroscopic quadrupole moment measured for the
3$^+_1$ ground state, $Q_{_S}(3^+_1)=0.08472(56)$ eb~\cite{nesbet1970atomic}, suggests a similarly dominant prolate shape arising from the $\alpha+d+\alpha$ cluster configuration,
in agreement with {\scshape AMD} density distributions~\cite{morita2016isospin}.
Inclusion of the octupole and/or hexadecapole degrees of freedom could enhance the lower region, bringing it into closer resemblance with Fig.~\ref{fig:20Ne};
however, the focus here is on the emergence of clusterization from minimal assumptions.

Moreover, Fig.~\ref{shape20Ne} illustrates the shape evolution of selected $N$$=$$Z$ self-conjugate nuclei heavier than $^{20}$Ne  in the $sd$-shell:
$^{32}$S ($\beta$$=$$0.32$~\cite{vermeer1982quadrupole}, $\gamma$$=$$22.4^{\circ}$~\cite{triaxiality2ElenaNico})
and $^{36}$Ar ($\beta$$=$$-0.20$~\cite{orce2021reorientation}, $\gamma$$=$$36.7^{\circ}$~\cite{triaxiality2ElenaNico}).
The bowling-pin shape in $^{20}$Ne gradually fades, evolving into a kiwi-like shape in $^{32}$S and a round-cushion-like shape in $^{36}$Ar~\cite{orce2021reorientation}.
This geometry for $^{36}$Ar is not random~\cite{maruhn2006alpha}, but naturally arises from symmetry considerations
as 9$\alpha$ particles arranged in a grid~\cite{sosin2016alpha}, in agreement with Nilsson-model calculations for the ground state of $^{36}$Ar~\cite{cseh2010hyperdeformed}.
For heavier nuclei, Eqs.~\eqref{eq:sh1} and \eqref{eq:sh2} generally exhibit similar features.

This overall similarity of nuclear shapes obtained using Eqs.~\eqref{eq:sh1} and \eqref{eq:sh2} with increasing nuclear size
may point to a deeper significance of triaxiality in nuclei. Indeed, many such nuclei are found to be quadrupole deformed and exhibit
substantial triaxial deformations between $\gamma\approx20^{\circ}-40^{\circ}$~\cite{triaxiality2ElenaNico,triaxialityElenaNico},
indicative of a smearing out of the nuclear density in the $x$$-$$y$ plane. This behaviour may arise from the
superposition of multiple intrinsic configurations that collectively form the nuclear state.

More specifically, the superposition principle in quantum mechanics underlies the sharper shape features of $^{10}$B and $^{20}$Ne arising from Eq.~\eqref{eq:sh1}, which
yields the nuclear shape in an originally arbitrary coordinate system aligned with the  principal axes of the ellipsoid.
However, the use of  empirical shape parameters $\beta$ and $\gamma$, extracted from electric-quadrupole matrix elements, effectively captures the
shell structure by averaging over many microscopic configurations, thereby yielding the most probable nuclear shape.
In contrast, the conventional alignment of the coordinate system leading to Eqs.~\eqref{eq:sh2} assumes a single intrinsic configuration with a smoothed-out shape.
That the nuclear shape is, in fact, a superposition of projected intrinsic states with different deformations is
elegantly demonstrated by the distinct contributions of intrinsic densities from fermionic molecular dynamics ({\scshape FMD}) basis states,
which yield strongly overlapping $\alpha$ clusters in the final eigenstates~\cite{Neff,feldmeier2017nuclear}.
A similar approach was originally developed by Wheeler in the resonating group model ({\scshape RGM})~\cite{wheeler1937mathematical},
in which the nuclear wave function is expressed as a linear combination of all possible cluster configurations~\cite{tang1978resonating}.

\section{Conclusions}

In conclusion, this work highlights how the quasi-molecular model with macroscopic observables $\beta$ and $\gamma$ ---  extracted from experimental
transitional and diagonal electric-quadrupole matrix elements --- provides direct insight into the complex  many-body dynamics and collective behaviour of nuclei.
Although Eq.~\eqref{eq:sh1} does not explicitly include nuclear-structure effects, these seem to be captured by experimental deformation parameters.
This approach is by no means a substitute for a more detailed microscopic description of nuclear clusters,
where the nuclear shape may be calculated from first principles, but it is nevertheless reassuring --- and somewhat intriguing ---
that similar conclusions emerge directly from what may be regarded as macroscopic manifestations of the superposition principle.

\section*{Acknowledgements}

The author {\scshape J.N.O.} would like to acknowledge insightful discussions with {\scshape CERN} colleagues involved in the White Paper on future {\scshape LIC} measurements at the {\scshape LHC},
as well as with Elena Lawrie at iThemba {\scshape LABS}.
Both authors acknowledge travel support from the South Africa–{\scshape CERN} Collaboration, funded by the South African Department of Science and Innovation and the National Research Foundation. \\

\bibliographystyle{iopart-num}

\bibliography{rise}

\end{document}